\documentclass[preprint,number,12pt]{elsarticle}

\journal{Astroparticle Physics}
\usepackage{graphicx}
\usepackage{amsmath}
\usepackage{txfonts}
\usepackage{natbib}

\newcommand{\bref}[1]{#1}
\newcommand{\breftwo}[1]{#1}

\newcommand{\eh}[1]{\,\mathrm{#1}}

\newcommand{\gev}{\eh{GeV}}

\newcommand{\dg}{^{\circ}}

\newcommand{\pct}{\eh{\%}}

\newcommand{\mr}[1]{\mathrm{#1}}

\renewcommand{\epsilon}{\varepsilon}

\newcommand{\tin}[1]{_{\mr{#1}}}

\newcommand{\w}{\omega}
\newcommand{\beq}{\begin{equation}}
\newcommand{\eeq}{\end{equation}}

\newcommand{\beqa}{\begin{eqnarray}}
\newcommand{\eeqa}{\end{eqnarray}}

\newcommand{\eref}[1]{Eq.\ \ref{#1}}

\newcommand{\sig}{\eh{\sigma}}

\begin{document}

\begin{frontmatter}

\title{A generalized likelihood ratio test statistic\\for Cherenkov telescope data}
\author{S. Klepser}
\address{IFAE, Edifici Cn., Campus UAB, E-08193 Bellaterra, Spain}
\ead{stefan.klepser@desy.de}

\begin{abstract}
Astrophysical sources of TeV gamma rays are usually established by Cherenkov
telescope observations. These counting type instruments have a field of view of few degrees
in diameter and record large numbers of particle air showers via their
Cherenkov radiation in the atmosphere. The showers are either induced by gamma
rays or diffuse cosmic ray background.
The commonly used test statistic to evaluate a
possible gamma-ray excess is Li and Ma (1983), Eq.\ 17, which can be
applied to independent on- and
off-source observations, or scenarios that can be approximated as such. This formula
however is unsuitable if the data are taken in so-called "wobble" mode
(pointing to several offset positions around the source), if at the same time the
acceptance shape is irregular or even depends on \bref{operating parameters
such as the pointing direction or telescope multiplicity}. To
provide a robust test statistic in such cases, this paper explores a possible
generalization of the likelihood ratio concept on which the formula of Li and
Ma is based. In doing so, \bref{the multi-pointing nature of the
data and the typically known instrument point spread function are fully
exploited} to derive a
new, semi-numerical test statistic.
Due to its flexibility and robustness against systematic uncertainties, it is not only useful for detection
purposes, but also for skymapping and source shape fitting. \bref{Simplified
Monte Carlo simulations are presented} to verify the results, and \bref{several
applications and further generalizations of the concept are discussed}.
\end{abstract}


\end{frontmatter}

\section{Introduction} \label{sec:intro}

The field of very-high-energy (VHE, $>100\gev$) gamma-ray astronomy is currently in its
third generation of instruments, with an advanced future project, CTA \citep{cta}, 
already being in its preparatory phase. The currently operated systems H.E.S.S.
\citep{hess}, MAGIC \citep{magicstereoperformance} and
VERITAS \citep{veritas} have established more than 100 VHE sources in the
sky\footnote{http://tevcat.uchicago.edu/}. 
The acceptence of these telescope systems covers few degrees in
diameter, and declines smoothly with increasing distance from the pointing
position. 
Establishing a gamma-ray signal from an astrophysical source requires to
significantly \bref{prove} a gamma-ray excess over
background events that typically remain dominant even after selection cuts. This background is
mostly composed of diffuse hadrons, part of which appears almost identical
\bref{to}
electromagnetic showers and has to be considered to be irreducible \citep{irreducible}. Besides
these "gamma-like hadrons", the irreducible background also contains smaller fractions of diffuse electrons and gamma rays. \bref{This irreducible background, and the statistical and systematic
uncertainties that come with it, are one of the main limiting factors of TeV astronomy; usually,
an observational campaign for a given source either reveals one source or
none, and
only in few cases, or if the effort of a large scan is undertaken, additional
unexpected sources are detected. Therefore, a statistical source detection
technique that is both sensitive to weak sources and stable against systematic
effects is of crucial importance to the field.}

The standard test
statistic to evaluate an excess of gamma rays from a given sky direction is
Li and Ma \cite{lima}, Eq.\ 17, which hereafter will be refered to as $S\tin{LM}$. It
was established among several alternatives they evaluated in their paper,
based on the fact that this likelihood ratio test statistic was the
only one that yielded a satisfyingly Gaussian null-hypothesis distribution.
The formula they presented was designed to compare the
event numbers of an on- and off-source observation ($N\tin{on}$,
$N\tin{off}$), and allows for a scaling factor $\alpha$
between the effective observation times ($t\tin{on} = \alpha\, t\tin{off}$) to account for unequal exposures.

\begin{figure}
\centering
\includegraphics[width=13cm]{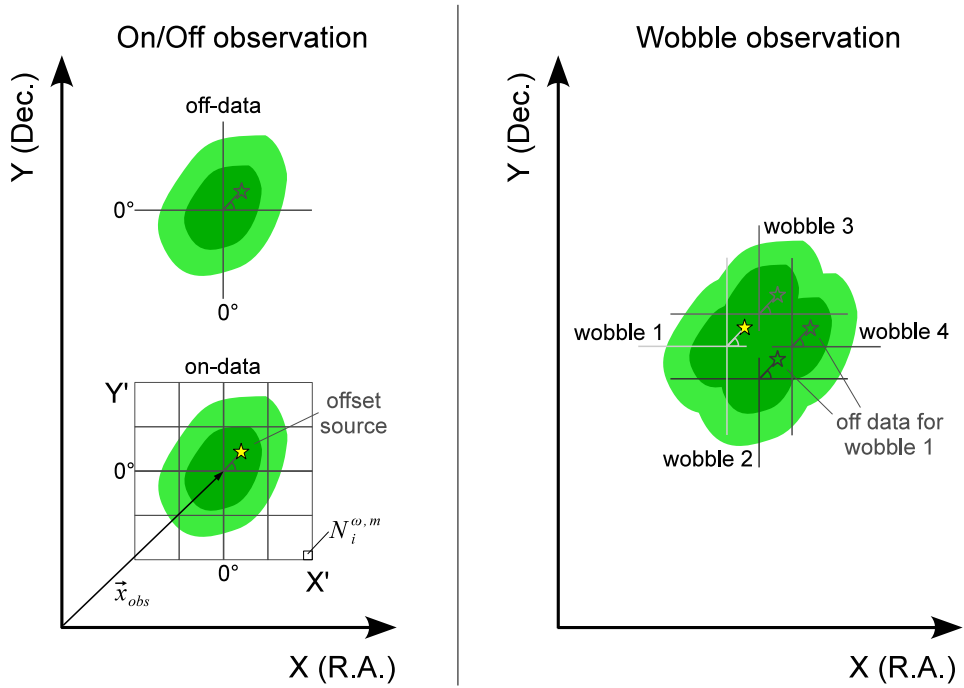}
\caption{
Observation schemes with asymmetric acceptance shapes (green areas). The sky
position to be evaluated for a signal is marked with a yellow star, the
corresponding off-regions with grey stars. Left: Original On/Off scheme. The off-data provides $N\tin{off}$,
the on-data $N\tin{on}$, and it needs one exposure ratio
$\alpha=t\tin{on}/t\tin{off}$ between those.
Right: Wobble scheme with four observation positions. The off data of each
wobble set can be
taken from all other wobble sets, resulting in four $\alpha$ parameters and
their corresponding $N\tin{on}$, $N\tin{off}$.
}
\label{fig:on_off_wobble}
\end{figure}

In modern observation practice, most Cherenkov telescope data are \textit{not} taken
in On/Off-mode (see Fig.~\ref{fig:on_off_wobble}, left), because this strategy implies a lot
of observation time dedicated to empty sky regions. Also, it is
prone to systematic
differences between the on- and off-data caused by instabilities in electronic
or atmospheric \bref{operating}
conditions, especially if the off-data \bref{could not be scheduled
contemporaneously enough with the on-source observation.}
Therefore, usually the "wobble" technique \citep{wobble} is applied, in
which the data  are taken at two or more observation positions offset in
different directions from the main
target coordinates (see Fig.~\ref{fig:on_off_wobble}, right). In this way, each
wobble set
provides both on- and off-data at the same time. In its original idea, the
exposure shape is considered to have some circular symmetry, and for each wobble set
the off-data can be taken from the same observation, at sky positions of
similar distance to the telescope pointing position
("reflected regions" \citep{reflectedregion}).
In that case, the off-exposure ratio $\alpha$ is the same for all wobble data
sets, and $S\tin{LM}$ can be applied after summing up the
on- and off-events of all wobble sets. This constant $\alpha$ can also be
achieved approximately if the off-data are taken from
\textit{hadron-like} background events
("template background", \citep{templatebackground}), or a ring area around the source position ("ring
background" \citep{hessskymapping}), both of which require \bref{also
symmetry assumptions or} an efficiency
correction through Monte Carlo simulations of the isotropic
background.

In the general case, though, the acceptance symmetry might not hold and a Monte
Carlo  correction may involve too high uncertainties.
This occurs for instance in very \bref{low-energy 
observations}, where camera-based acceptance inhomogeneities (dead pixels,
trigger fluctuations) are both difficult
to model in simulations and furthermore lead to \bref{features in the
acceptance shape that can depend on the Alt/Az pointing direction or other operation parameters
of the system}. This is particulary troublesome in two-telescope systems like MAGIC
\citep{magicstereoperformance}, where already the basic geometrical overlap of the two
fields of view implies an elongated, Alt/Az-dependent exposure \bref{shape}.
Under these conditions, the "reflected regions"
approach does not hold, and the 
off-data for each wobble set have to be taken
from the other wobble sets (see Fig.~\ref{fig:on_off_wobble}, right). This
approach can provide \bref{a sensitive measurement}, because with several wobble
sets, the off regions are numerous and well-populated, but it results in a
different $\alpha$ for each wobble set, which is not supported by $S\tin{LM}$.
On top of that, if this
procedure has to be done \bref{separately for different types of data (be it
for instance different Azimuth angles or telescope multiplicities)}, it leads to many more
$\alpha$-parameters, and in general, some off-events may happen to
be oversampled if they lie in more than one off-region, which is also not
considered in $S\tin{LM}$.
As a consequence, while the 
background density can still be modeled under certain assumptions and some
 numerical effort \citep{icrcadvanced}, the
test statistic $S\tin{LM}$, if still applied in some way, becomes very
approximative. This may be dealt with in practice
by making additional high requirements to the signal-to-background ratio of a
detection \citep{magicstereoperformance}, which however \bref{limits the
effective sensitivity of the instrument}.

Therefore, unlike previous efforts \citep{hessskymapping,icrcadvanced}, this work will not pursue to extract the
variables needed for $S\tin{LM}$ through the
complex task of explicit background modeling. Instead, \bref{the
likelihood ratio concept behind that test statistic will be generalized, and new
formulae will be derived} that can directly be applied to
multi-wobble Cherenkov telescope data. To do so, \breftwo{no} Monte
Carlo simulations or exposure symmetry assumptions will be required. It will
only be assumed that
the telescope acceptance shape is the same for different wobble data sets if
they are taken \bref{under similar operating conditions}\footnote{Note that if this basic assumption does not
hold, any other symmetry assumption also breaks.}.

Besides that, \bref{this work will also address the disadvantage of $S\tin{LM}$
that it} depends on the size of the
signal region in which $N\tin{on}$ is calculated. This area is
usually defined through an integration radius ("$\theta^2$-cut", with
$\theta$ being the angular distance between reconstructed gamma direction and
source position). Its optimization either depends on the background density
and an assumption of the source
strength, or may involve several trials. \bref{In this paper,} these
assumptions \bref{are reduced} by
accomodating the knowledge of the point spread function (PSF) in the formulae,
which makes them independent of the source strength or background density.

Likelihood ratios are frequently used to
convert a complex likelihood maximization problem to a test statistic that
follows a $\chi^2$-distribution. This possibility was first proven in
\cite{llhratio}, and was suggested for astronomical purposes in \citep{llhratioastronomy}. The technique is now widely used in counting type
experiments, mainly in X-ray and gamma-ray astronomy, and can be
applied both for detection and
optimization purposes \citep{llhrvschi2}. \bref{It should be pointed out} that criticism
and potentially more accurate or more general alternatives
to the likelihood ratio concept exist \citep{bayesianlima,limacritics,fisherexactlima}, but are not subject of this work.

The structure of the paper is to first define the coordinate systems and
naming conventions needed for the calculations. Then, a likelihood function is set up and maximized to
gain all relevant parameters \breftwo{(Sec.~\ref{sec:llh})}. Based on that,
the test statistic is formulated in Sec.~\ref{sec:ts}
and its intrinsic inclusion of $S\tin{LM}$ is demonstrated. In
Sec.~\ref{sec:generalizations}, some further possible generalizations and
applications of the formulae are
discussed and a recipe for "Likelihood Ratio Skymapping" is suggested. Finally, some
example toy simulations are shown in Sec.~\ref{sec:mc} to verify the method.

\section{Building the likelihood function} \label{sec:llh}

In this section, a binned likelihood function is formulated which will be the basis
of the likelihood ratio test statistic, but can also serve to fit the shape or
positional parameters of a detected source. It is set up as a Poissonian
\bref{probability
function} that evaluates the consistency of the different wobble
subsets \bref{taken under a given operating condition} with each other, allowing for a hypothetical
source with a well-defined shape. It is a binned likelihood, but can be
generalized to an unbinned likelihood easily (Sec.~\ref{subsec:unbinned}).

\bref{In this paper, the term "operating condition" may in practice refer to a
range in any quantity that influences the acceptance shape. This binning may for
instance be done in the telescope pointing direction (Alt/Az) for two-telescope
systems, or in the telescope multiplicity for a multi-telescope system like CTA.
It may also
be binned in atmospheric conditions, night sky background light level or
discrete performance states of the camera. 
Of course, the formulae are equally valid if no such binning needs to be
applied.}

\subsection{Namings and definitions} \label{subsec:namings} 

The data are assumed to be taken as several wobble sets $\w=1,2,\dots,W$,
centered at sky coordinates $\mathbf{x}_{\mr{obs},\w}$ ("pointing directions"), throughout 
\bref{operating conditions}
$m=1,2,\dots,M$ \bref{(see previous paragraph)}. Hence, the observation can be described as a set of $W\times M$
two-dimensional sky histograms. These histograms shall be set up in
\textit{relative} sky coordinates ($\mathbf{x'} =
\mathbf{x}-\mathbf{x}_{\mr{obs},\w}$)
centered at the observation direction $\mathbf{x}_{\mr{obs},\w}$ of each
wobble set (see grid on the "on-data" drawing of
Fig.~\ref{fig:on_off_wobble}). The individual relative sky bins \bref{are} named
$i = 1,2,\dots,I$, \bref{where} $i$ may be regarded as a
representation of a two-dimensional bin $(i_x,i_y)$ without loss of
generality. The number of gamma-like events in one such bin
$(\w,m,i)$ is $N^{\w,m}_i$.
In relative sky coordinates,
the shape of the background event distribution is the same for all histograms
\bref{that belong to a given operating condition}, while a signal at a fixed absolute sky position will
appear in different relative locations for different wobble sets.
In the following, the axes of the relative sky histograms ($X'$, $Y'$ in
Fig.~\ref{fig:on_off_wobble}) are treated as synonyms for (relative) right ascension and
declination, although in practice, one should replace those for axes that are
truly rectangular throughout the field of view.

The exposure of \bref{the data taken under a given operating condition} $m$ is
distributed among the wobble sets $\w$ in ratios of $a_{\w}^m$, such that
$\sum_{\w}a_{\w}^m = 1$. There are different technical solutions of how to
evaluate these ratios, the simplest and most common being through the observation
times
\beq
a_{\w}^m = \frac{t_{\w}^m}{t\tin{tot}^m}.
\eeq
which is correct if the background rate is the same in all wobble sets of $m$.
In the absence of a signal, the expectation value $E(N^{\w,m}_i)$ for a given
bin ($\w,m,i$) equals the background expectation
$n^{\w,m}_i$, which in turn is a well-defined fraction of
the total background exposure \bref{taken under the operating condition} $m$:
\beq 
E(N^{\w,m}_i) = n^{\w,m}_i = a_{\w}^m\, n^{m}_i.
\eeq
If a signal is present, its shape can be modeled with a (normalized) kernel
$g^{\w}_i$, which usually may be a function representing the \bref{gamma-ray} PSF of the instrument, or a more extended shape for dedicated searches
of extended sources. \bref{The gamma-ray PSF is usually a well-known
performance characteristic that is determined from
simulations and/or observations of strong known point sources. It might, as
any other data cut, slightly depend on the assumed spectral index of the
source.}
The kernel parameters $g^{\w}_i$ depend on $\w$  because
the relative source coordinates depend on the wobble offset. Also, the shape
of the PSF (i.e. the resolution) might depend on the distance from the
pointing direction.

\bref{The kernel, which only characterizes the shape of the signal, is now
\breftwo{multiplied by a scaling} factor $\phi$ to constitute a signal term which is added} to the expectation value as follows:
\beq \label{eq:expectation}
E(N^{\w,m}_i\,|\,\phi) = a_{\w}^m\, n^{m}_i\,(1\,+\,\phi\,g^{\w}_i)
\eeq
This way, the excess events implied by a given \textit{relative
\breftwo{excess}} $\phi$
of a source is automatically proportional to the efficiency of the detector at
the position $i$ \bref{in the operating condition} $m$, and \bref{to} the
exposure of the given wobble set $\w$. This first-order
approximation assumes that the
background exposure $n^{\w,m}_i$, which is proportional to the efficiency to
gamma-like hadrons, is also proportional to the \bref{gamma-ray} efficiency (see
Sec.~\ref{subsec:ghgeneral} for a way to loosen
this assumption).

\breftwo{The expectation value of the relative excess $\phi$ from a gamma-ray
source scales with its absolute flux. Therefore, estimators for $\phi$ can be
used for flux skymapping (after the efficiency correction
outlined in Sec.~\ref{subsec:ghgeneral}). Still, the meaning of $\phi$ is technically
that of an excess (or deficit), and therefore it is not, like a flux, bound to be
positive or zero. Also, its interpretation as a gamma-ray
flux is not exclusive --- a significantly non-zero
excess can also be caused by other physical effects as for instance the moon
shadow, or systematic detector artefacts not common to all
wobble datasets.
In any
of these cases,
the null hypothesis for the excess/deficit judgement is $\phi=0$, which is a degenerate case of the
signal hypothesis, and not at the edge of the parameter space. This is
important for the following, because like this, the likelihood ratio test
statistic can be expected to follow a $\chi^2$-distribution. A remaining
limitation of the $\phi$ parameter space is the fact that the expectation value for the event number,
$E(N^{\w,m}_i\,|\,\phi)$, has to be positive, so $\phi$ has to be larger than
$-1/g^{\w}\tin{max}$.}

While the normalization of the kernels $g^{\w}_i$ has to be consistent
among different $\w$, it neither has to be unity, nor does $\sum_i g^{\w}_i$
have to be the same for different $\w$. For example, if additional
off-observations are added to the analysis, they would have $g^{\w}_i=0$ for
all $i$. Also, Sec.~\ref{subsec:ghgeneral} discusses a possible case where a varying
normalization of $g^{\w}_i$ is appropriate.

For convenience in the following calculations, also the
\textit{averaged kernel} for a given \bref{operating condition} is defined:
\beq  \label{eq:avkern}
g^{m}_{i} \equiv \sum_{\w} a_{\w}^{m}\,g^{\w}_{i}
\eeq
and the sum of events in a given bin $(m,i)$:
\beq \label{eq:nullexp}
N^{m}_{i} \equiv \sum_{\w} N^{\w,m}_{i}
\eeq

\subsection{The likelihood function}

The \bref{likelihood function can be defined as the product of all Poissonian
probability functions of the bins in $i$, $\w$ and $m$:}
\begin{align}
L(N^{\w,m}_i\,|\,n^{m}_i,\phi) & =  \prod_m \prod_{i} \prod_{\w} \frac{(a_{\w}^m\,n^{m}_i\,(1\,+\,\phi\,g^{\w}_i))^{N^{\w,m}_i}}{N^{\w,m}_i!}\,e^{-a_{\w}^m\,n^{m}_i\,(1\,+\,\phi\,g^{\w}_i)} \\
   & =  \prod_m \prod_{i} e^{-n^{m}_i\,(1\,+\,\phi\,g^{m}_i)} \prod_{\w} \frac{(a_{\w}^m\,n^{m}_i\,(1\,+\,\phi\,g^{\w}_i))^{N^{\w,m}_i}}{N^{\w,m}_i!} 
\end{align}
The latter step took advantage of the normalization of
$a_{\w}^m$ within each $m$ and the average kernel convention (\eref{eq:avkern}).

For the maximization procedure, it is convenient to calculate the
log-likelihood $\mathcal{L} = \ln L$, which is
\beq \label{eq:lllhf}
\mathcal{L}(N^{\w,m}_i\,|\,n^{m}_i,\phi)  = K + \sum_m \sum_i \left[
-n^{m}_i\,(1\,+\,\phi\,g^{m}_i) + \sum_{\w}
N^{\w,m}_i\,\ln\left(n^{m}_i\,(1\,+\,\phi\,g^{\w}_i)\right)\right]
\eeq
All terms that are independent of the free parameters $n^{m}_i,\phi$ were
absorbed into the constant $K$.

\subsection{Determination of the parameters}

The free parameters to be optimized in order to maximise the likelihood
\bref{function} are $I\times M$ total exposures $n^{m}_i$ of the \bref{operating conditions}, and, if a signal at a given sky
position is assumed, its corresponding $\phi$.
The following paragraphs show that this is possible analytically
for the $n^{m}_{i}$, and numerically for
the \breftwo{relative excess} parameter $\phi$.

\subsubsection{Background density parameters $n^{m}_{i}$}

To find the $n^{m}_{i,\phi}$ that maximize the likelihood \bref{function} for a given $\phi$,
\bref{one can}
calculate the first partial
derivative of the log-likelihood function (\eref{eq:lllhf}) for a given $m'$
and $i'$:
\begin{equation}
\begin{split}
\frac{\partial \mathcal{L}(N^{\w,m}_{i'}\,|\,n^{m}_{i},\phi)}{\partial n^{m'}_{i'}}
 \quad & = \quad -1 - \phi\,g^{\w}_{i'} + \sum_{\w} \frac{N^{\w,m'}_{i'}}{n^{m'}_{i'}}
\\
 \quad & =  \quad -1 - \phi\,g^{\w}_{i'} + \frac{1}{n^{m'}_{i'}}\,N^{m'}_{i'}
\end{split}
\end{equation}
This expression equals zero if
\beq \label{eq:optexp}
  n^{m}_{i,\phi} = \frac{N^{m}_{i}}{1+\phi\,g^{m}_{i}}.
\eeq
The second derivative of the log-likelihood function is always negative for
this solution, so it is the maximum of the likelihood function for a given
$\phi$. Note also
that for the null hypothesis, $\phi=0$, the best approximator is,
intuitively,
\beq
n^{m}_{i,0} = N^{m}_{i}.
\eeq

\subsubsection{Relative excess parameter $\phi$} \label{subsec:phiopt}

Inserting the optimized exposures (\eref{eq:optexp}) into the log-likelihood function (\eref{eq:lllhf}),
the log-likelihood expression for a given $\phi$ parameter can be simplified:
\beqa
\mathcal{L}(N^{\w,m}_i\,|\,\phi)
& = &   K' + \sum_m \sum_i \left[ -N^{m}_{i}  + \sum_{\w}
N^{\w,m}_i\,\ln\left(N^{m}_{i}\,\frac{1\,+\,\phi\,g^{\w}_i}{1+\phi\,g^{m}_{i}}\right)
\right]\\ 
 & = & K'' + \sum_m \sum_i \sum_{\w}
N^{\w,m}_i\,\ln\left(\frac{1\,+\,\phi\,g^{\w}_i}{1+\phi\,g^{m}_{i}} \right)
 \label{eq:lllhfphi}
\eeqa
Again, $K'$ and $K''$ represent terms that are independent of $\phi$.
 
The partial derivative w.r.t.\ $\phi$ is
\beq \label{eq:lllderphi}
\frac{\partial \mathcal{L}(N^{\w,m}_i\,|\,\phi)}{\partial \phi}
  = \sum_m \sum_i \sum_{\w}
N^{\w,m}_i\,\frac{g^{\w}_i-g^{m}_{i}}{(1+\phi\,g^{\w}_i)(1+\phi\,g^{m}_{i})}
\eeq
As it turns out in simulations, finding the root of this expression, and
verifying the positive second derivative, is numerically straight-forward. In all reasonable cases, it leads to one solution $\phi\tin{sup}$ that maximises the
\bref{likelihood} function (see
Sec.~\ref{sec:caveats} for possible exceptions). 

\section{The likelihood ratio test statistic} \label{sec:ts}

\subsection{The likelihood ratio}

The likelihood ratio is defined as the ratio of the maximal \bref{likelihood} of the null
hypothesis ($\phi=0$), and the global maximum of the \bref{likelihood}, allowing for a
signal ($\phi\not= 0$):
\begin{equation} \label{eq:llhr}
\begin{split}
\Lambda \quad  = &\quad \frac{L(N^{\w,m}_i\,|\,n^{m}_{i,0},\phi=0)}{L(N^{\w,m}_{i,\phi}\,|\,n^{m}_{i, \phi},\phi)} \\
         = &\quad \prod_m \prod_{i} e^{n^{m}_{i,\phi}\,(1\,+\,\phi\,g^{m}_i)-n^{m}_{i,0}} \prod_{\w} \left(\frac{n^{m}_{i,0}}{n^{m}_{i,\phi}\,(1\,+\,\phi\,g^{\w}_i)}\right)^{N^{\w,m}_i}\\
         = &\quad \prod_m \prod_{i} \prod_{\w} \left(\frac{1\,+\,\phi\tin{sup}\,g^{m}_i}{1\,+\,\phi\tin{sup}\,g^{\w}_i}\right)^{N^{\w,m}_i}
\end{split}
\end{equation}
The typical prescription to convert this to a test statistic \citep{llhratio}
is $\mr{TS} = -2 \ln \Lambda$.
Since the null hypothesis has $M\times I$ parameters, and is a special case of
the alternative hypothesis, which has $M\times I + 1$ parameters (because of
$\phi$), the difference in degrees of freedom is $1$, and $\mr{TS}$
can be expected to follow a $\chi_1^2$ distribution. \bref{As in the case of
$S\tin{LM}$, this only holds for high count numbers; in fact, since the
underlying mathematics are the same, it can be assumed
that the validity into the low-count regime is similar to that of $S\tin{LM}$}.

\bref{Since $\sqrt{\mr{TS}}=\sqrt{\chi_1^2}$ is a half-normal distribution,}
\beq \label{eq:significance}
S = \sqrt{\mr{TS}} = \sqrt{2 \times \sum_m \sum_{i} \sum_{\w} N^{\w,m}_i
\,\ln\left(\frac{1\,+\,\phi\tin{sup}\,g^{\w}_i}{1\,+\,\phi\tin{sup}\,g^{m}_i}\right)}
\eeq
is the Gaussian significance of the considered sky position to contain a
gamma-ray excess \bref{(or deficit)}. This holds independently of whether the initial
assumptions about the gamma efficiency \bref{and PSF are correct, since
imprecise assumptions cannot contradict the null hypothesis and might
therefore only
make the test statistic less sensitive, but never wrong.} Section~\ref{subsec:gauslima} shows
a validation of this Gaussian nature in simulation.

While the test statistic may be regarded 
semi-numerical due to the determination of $\phi\tin{sup}$, it is unambiguous
and can be determined at any desired precision. The trials usually
needed to choose the optimal radius of the signal region are obsolete, because
the PSF of the instrument is incorporporated, which is typically
well-known \bref{(see sec.~\ref{subsec:namings})}.

For a given signal $\phi\tin{sup}$, the significance after
\eref{eq:significance} depends on how much $g^{\w}_i$ and $g^{m}_i$ differ.
Therefore, a higher number of wobble sets quite naturally leads to a higher
\bref{significance}
without increasing the complexity of the analysis by manual selection of valid off-region(s).
Furthermore, the
formulae are invariant against the number of \bref{different operating conditions} $M$, as long as the
coverage of the wobble sets is good enough to provide well-balanced exposure ratios
$a_{\w}^m$ needed for the $g^{\w,m}_i$ to differ.

\subsection{The degenerate case of on-off-analysis} \label{sec:lima}

In this section, instead of the general multi-wobble case with
\bref{variable exposure shape} and an arbitrary PSF, \bref{a
case of only one \bref{operating condition} ($M=1$) is considered, with} separate on- and off-target
observations ($W=2$), and a simple step function kernel that represents a
predefined on-target sky region. The observation times are
$t\tin{on} = \alpha\, t\tin{off}$, so the above $a_{\w}$ relate
to the $\alpha$ as follows:
\beqa
a_1 = \frac{\alpha}{\alpha + 1}\\
a_2 = \frac{1}{\alpha + 1}
\eeqa
Here and in the rest of this section, the index $1$ refers to the on-observation and $2$ to the
off-observation. The step function kernel is 1 within an arbitrarily shaped signal region
($i\in\mr{SR}$), and zero elsewhere.
Consequently, the kernel constants are
\beqa
g^{\w}_i & = &
\begin{cases}
1 & \text{if } \w = 1 \text{ and } i \in\mr{SR} \\
0 & \text{else} 
\end{cases}\\
g^{m}_i  & = &
\begin{cases}
\frac{\alpha}{\alpha+1} & \mbox{if } i \in\mr{SR} \\
0 & \mbox{else}.
\end{cases}
\eeqa
Inserting these terms in \eref{eq:lllderphi}, it can be converted to
\beqa
\frac{\partial \mathcal{L}(N^{\w,m}_i\,|\,\phi)}{\partial \phi}
  & = & \frac{\sum_{i\in\mr{SR}} N^{\w=1}_i}{(\alpha+1)(\phi+1)(\phi\frac{\alpha}{\alpha+1}+1)} +
    \frac{-\alpha\, \sum_{i\in\mr{SR}}N^{\w=2}_i}{(\alpha+1)(\phi\frac{\alpha}{\alpha+1}+1)} \\
  & = & \frac{N\tin{on}}{(\alpha+1)(\phi+1)(\phi\frac{\alpha}{\alpha+1}+1)} -
    \frac{\alpha N\tin{off}}{(\alpha+1)(\phi\frac{\alpha}{\alpha+1}+1)}
\eeqa
the root of which can analytically be calculated to be
\beq
\phi = \frac{N\tin{on}-\alpha N\tin{off}}{\alpha N\tin{off}}.
\eeq
Putting this to \eref{eq:significance}, \bref{one finds}
\beq
S = \sqrt{2 \left[N\tin{on}
\ln\left(\frac{1+\alpha}{\alpha}\frac{N\tin{on}}{N\tin{on}+N\tin{off}}\right) + N\tin{off}
\ln\left((1+\alpha)\frac{N\tin{off}}{N\tin{on}+N\tin{off}}\right)\right]}
\eeq
which is the well-known Eq.\ 17 of Li and Ma \cite{lima}, making it a degenerate case of
\eref{eq:significance} of this work. The advantage of the latter is that it combines an
arbitrary amount of differently populated wobble (or off-target) observations,
and at the same time
uses the actual PSF of the instrument instead of a discrete on-target area.
This also avoids the need for an ambiguous choice of integration radius.

\section{Further generalizations and suggested
applications}\label{sec:generalizations}

\subsection{Establishing several sources and their parameters} \label{sec:multisource}

Once a source is detected, the log-likelihood function (\eref{eq:lllhfphi}) is fully adequate
to estimate the parameters of the source, such as the position or extension.
These source parameters can be
regarded as parameters of the kernel $g^{\w}_i$ (which the constant $K''$ does not
depend on).

If one or several sources could be established \breftwo{with relative excesses} $\phi_n$,
and corresponding kernel parameters $g^{\w}_{i,n}$, $g^{m}_{i,n}$,
they can be inserted to \eref{eq:expectation} in order to scan the field for
other sources, and cross-check the new null-hypothesis distribution:
\beq \label{eq:expectation2}
E(N^{\w,m}_i\,|\,\phi) = a_{\w}^m\, n^{m}_i\,(1\,+\,\phi\,g^{\w}_i\,+\,\sum_n
\phi_n\,g^{\w}_{i,n})
\eeq
In this case, the log-likelihood derivative (\eref{eq:lllderphi}) turns to
\beq \label{eq:lllderphi2}
\frac{\partial \mathcal{L}(N^{\w,m}_i\,|\,\phi)}{\partial \phi}
  = \sum_m \sum_i \sum_{\w}
N^{\w,m}_i\,\frac{g^{\w}_i(1+\sum_n \phi_n\,g^{m}_{i,n})-g^{m}_{i}(1+\sum_n
\phi_n\,g^{\w}_{i,n})}
{(1+\phi\,g^{\w}_i+\sum_n \phi_n\,g^{\w}_{i,n})
 (1+\phi\,g^{m}_{i}+\sum_n \phi_n\,g^{m}_{i,n})},
\eeq
and also the test statistic derived from the likelihood ratio is now different, because
the null hypothesis already assumes the presence of sources:
\beq \label{eq:significance2}
S = \sqrt{\mr{TS}} = \sqrt{2 \times \sum_m \sum_{i} \sum_{\w} N^{\w,m}_i
\,\ln\left(\frac{1\,+\,\frac{\phi\tin{sup}\,g^{\w}_i}{1\,+\,\sum_n \phi_n\,g^{\w}_{i,n}}}
                {1\,+\,\frac{\phi\tin{sup}\,g^{m}_i }{1\,+\,\sum_n \phi_n\,g^{m}_{i,n}}}\right)}
\eeq
Using these modified formulae is correct as long as the sources and their
effective off-regions are
\bref{spatially} independent of each other. If not, then their averaged kernels
$g^{m}_{i,n}$ overlap, and the $\phi_n$
cannot be optimized independently, in which case a multi-parameter maximization, or an
iterative scheme has to be applied.

\subsection{Likelihood Ratio Skymapping} \label{sec:procedure}

A common task with Cherenkov telescope data is to scan the
whole field of view for unknown sources. Since the formulae provided here are
perfectly viable to do this even in the general case of an unknown,
\bref{operating-condition}-dependent acceptance shape, \bref{I} suggest the following "Likelihood Ratio
Skymapping" procedure:

\begin{enumerate}
\item The events are filled into histograms $N^{\w,m}_i$, and the relative
exposure ratios $a_{\w}^m$ \bref{are} determined. 
\item A grid of absolute sky positions is defined. The bins are independent of those of the
wobble histograms and should be large enough to avoid oversampling (see
Sec.~\ref{sec:caveats}).
\item For each grid point, the following procedure is applied:
  \begin{enumerate}
  \item The kernel constants $g^{\w}_i$ are calculated for each wobble set
$\w$ in relative sky coordinates.
  \item The average kernels $g^{m}_i$ are calculated, using the weights
$a_{\w}^m$ (\eref{eq:avkern}).
  \item The root of \eref{eq:lllderphi} is determined to obtain $\phi\tin{sup}$.
  \item The significance is calculated after \eref{eq:significance}, using the
sign of $\phi\tin{sup}$ to distinguish positive from negative excess.
  \end{enumerate}
\item If a source is found, it is modeled with a log-likelihood fit
(\eref{eq:lllhfphi}), and the whole procedure is repeated with the
modified formulae of \ref{sec:multisource} until the significance distribution
follows the null-hypothesis distribution of a Gaussian function.
\end{enumerate}

In this scheme, which follows a similar iterative strategy as the likelihood analysis
of the Fermi Science
Tools\footnote{http://fermi.gsfc.nasa.gov/ssc/data/analysis}, the final result is a number of established sources, an
empty significance skymap, and a significance distribution that meets the null hypothesis.
There are several ways one might display these sources in a publication while
avoiding the negative excesses elsewhere, but \breftwo{in this work, no} fixed
recommendation concerning this graphical issue will be pursued.

\subsubsection{Technical notes and caveats}\label{sec:caveats}

Calculating and interpreting a skymap requires some more considerations than
the mere calculation of a significance value. Some of these concern features
that can be regarded as caveats of the method introduced here, and some are features that equally occur in other VHE skymapping methods.

\begin{enumerate}
\item Features common to most skymapping methods
\begin{enumerate}
 \item In very poorly populated
 data sets, it can happen that the log-likelihood derivative
 (\eref{eq:lllderphi}) does not approach zero (in empty
 sky regions), or the likelihood increases towards negative $\phi$ until it
 meets its limit $-1/g^{\w}\tin{max}$. Both are caveats that equally apply to
 other skymapping methods, because the \breftwo{negative excess} that can be considered is
 always limited by the fact that no negative event numbers can be measured. In
 Sec.~\ref{sec:mc}, satisfying results are obtained by discarding skybins that belong
 to the former case (in areas without sensitivity), and using the limit of $S$
 when $\phi$ approaches $-1/g^{\w}\tin{max}$.
 \item If the grid of the skymap is small compared to the
 kernel/PSF of the instrument, the single significance values $S$ become
 correlated and their distribution cannot be easily tested to be compatible
 with a Gaussian or not. This oversampling problem is also common to most
 skymapping methods and needs to be addressed with an appropriate binning or
trial strategy (see Sec.~\ref{sec:mc}), or an appropriate consideration of the
correlations in the compatibility test.
 \item In case of multiple sources (or, equivalently, very extended sources),
signal photons of one source might appear in the effective off regions of another, weakening the
significances of both. This issue can be addressed by observing in
multiple wobble sets, and a sufficient sky coverage around a potentially extended
source.
\end{enumerate}
\item Features specific to Likelihood Ratio Skymapping: 
\begin{enumerate}
 \item Since fewer assumptions go into the test statistic than in conventional
background estimation techniques \citep{hessskymapping}, it is much less affected
 by systematic uncertainties. However, it is
 also bound to be statistically somewhat less sensitive, at least in the two-wobble
case where each of the two data sets only contributes one off region. If many wobble
coordinates are used, every data set provides $W-1$ off regions, and
the reduced \bref{significance} can be expected to be compensated rapidly.
 \item \bref{As $S\tin{LM}$, also the formulae presented here} do not distinguish 
 between excess and deficit of events. Therefore, the presence of a signal produces a
 positive $S$ peak at the source position, but also a negative $S$ peak at
 the sky coordinates that the corresponding off-data are taken from. This effect cannot provoke false detections,
 and is compensated by the modified formulae in Sec.~\ref{sec:multisource}.
 Also, it looses importance with increasing number of wobble sets.
\end{enumerate}
\end{enumerate}

\subsection{Excess events}

If a number of excess events has to be calculated, this can be done
summing up the signal part of \eref{eq:expectation}:
\beqa 
N\tin{ex} & = & \sum_m \sum_i \sum_{\w} a_{\w}^m\, n^{m}_i\,\phi\tin{sup}\,g^{\w}_i \\
          & = & \sum_m \sum_i N^{m}_i\,\frac{\phi\tin{sup}\,g^{m}_{i}}{1+\phi\tin{sup}\,g^{m}_{i}} \label{eq:nex}
\eeqa
In case the signal of other sources has to be considered, as discussed in
Sec.~\ref{sec:multisource}, the corresponding expression is
\beq
N\tin{ex} = \sum_m \sum_i
N^{m}_i\,\frac{\phi\tin{sup}\,g^{m}_{i}}{1+\phi\tin{sup}\,g^{m}_{i}+\sum_n
\phi_n\,g^{m}_{i,n}}.
\eeq

\subsection{Variable gamma/hadron acceptance ratio, flux skymapping and
variable sources}\label{subsec:ghgeneral}

The relative \breftwo{excess} parameter $\phi$ is based on the assumption that
the gamma-ray acceptance is proportional to the acceptance for gamma-like
hadrons. However, Monte Carlo simulations may indicate that this is not the case, and
provide correction factors
\beq
\gamma_i^m = \frac{\epsilon_i^m(\mr{gammas})}{\epsilon_i^m(\mr{hadrons})},
\eeq
where the $\epsilon$ are the efficiencies to gammas and hadrons. In \bref{the
scheme presented here,} these constants may simply be incorporated into the kernel constants:
\beq
g^{\w}_i \rightarrow \gamma_i^m\,g^{\w}_i
\eeq
This way, or if $\gamma_i^m = 1$ is a good approximation, the factor between
hypothetical source flux and relative \breftwo{excess} parameters
$\phi\tin{sup}$ will be constant
throughout the field of view, and $\phi\tin{sup}$ may be used to
display the physically more relevant \breftwo{\textit{relative flux} skymap (possibly
omitting unphysical negative flux values for consistency)}. \bref{One has to
bear in mind though that $\phi\tin{sup}$ is a value that is relative to the
background density, so it is only reliable if the background density is
sufficiently high. In poorly populated skymaps, the number of excess
events (\eref{eq:nex}) may be a more stable parameter.}

Another considerable case where the kernel normalization may be modified is when the signal
is not constant in time, but variable, and an a-priori assumption of the light curve
is available.

\subsection{Unbinned analysis and "orbit" mode} \label{subsec:unbinned}

The presented formulae can be adopted for unbinned analysis if the
limit of an infinitely fine sky-binning $I$ is assumed. In this case, most bins are empty
($N^{\w,m}_i = 0$) and the sums in Equations
\ref{eq:lllhfphi} ,\ref{eq:lllderphi} and \ref{eq:significance} turn into sums over the relative
coordinates $\mathbf{x'_j}$ of the single events
$j = 1,2,\dots,J$ (i.e. bins for which $N^{\w,m}_i = 1$). Instead of the previously discrete kernel
constants, functions $g^{\w}(\mathbf{x'_j})$ have to be used.  In this case, the log-likelihood
function is
\beq
\mathcal{L}(\mathbf{x'_j}\,|\,\phi) = K'' + \sum_j
\ln\left(\frac{1+\phi\,g^{\w_j}(\mathbf{x'_j})}{1+\phi\,g^{m_j}(\mathbf{x'_j})}
\right),
\eeq
the derivative w.r.t\ $\phi$ is
\beq
\frac{\partial \mathcal{L}(\mathbf{x'_j}\,|\,\phi)}{\partial \phi} = \sum_{j}
\frac{g^{\w_j}(\mathbf{x'_j})-g^{m_j}(\mathbf{x'_j})}{(1+\phi\,g^{\w_j}(\mathbf{x'_j}))(1+\phi\,g^{m_j}(\mathbf{x'_j}))},
\eeq
and the test statistic is
\beq
S = \sqrt{2 \times \sum_{j}
\ln\left(\frac{1\,+\,\phi\tin{sup}\,g^{\w_j}(\mathbf{x'_j})}{1\,+\,\phi\tin{sup}\,g^{m_j}(\mathbf{x'_j})}\right)}.
\eeq
Although potentially more precise, this unbinned quantities may be
computationally much more expensive, since the function $g^{\w}(\mathbf{x'})$
has to be calculated $J\times W$ times for each evaluation of $\mathcal{L}$,
whereas for a binned analysis, only $I\times W$ calculations are needed, which
are typically much fewer calls.

A case where the unbinned approach might be computationally more effective
than the binned approach is the application
to "orbit mode" observations \citep{orbitmode}, in which the telescope pointing is not discrete,
but moves along a circle around the target coordinates. If one introduces the
\textit{wobble angle} $\xi$ that parametrizes the pointing direction along
that circle, the above equations still hold if $g^{\w_j}(\mathbf{x'_j})$ is
replaced by a variable kernel $g(\mathbf{x'_j}, \xi)$, and
$g^{m_j}(\mathbf{x'_j})$ is the convolution of $g(\mathbf{x'_j}, \xi)$ with the
distribution of $\xi$ within the \bref{operating condition} $m_j$ that the event $j$ belongs
to.

\section{Monte Carlo Simulations} \label{sec:mc}

To verify the proposed Likelihood Ratio Skymapping scheme and its practical
feasibility, \bref{two standard scenarios were simulated} for which the method might be useful. For
simplicity, again arbitrary
X-Y-coordinates \breftwo{are used} instead of right ascension and declination. The
histograms $N^{\w,m}_i$ are $3\dg\times 3\dg$ and the kernel (PSF) is assumed to be
a Gaussian function with $\sigma_{39\pct}=0.05\dg$. The bin sizes are $1\,\sigma_{39\pct}$
(although they may in principle be arbitrarily small, at expense of
computational load). The sky grid is conservatively chosen to be very coarse
($\sqrt{2\pi}\,\sigma_{39\pct}$)\footnote{The effective sky area covered by a
two-dimensional Gaussian kernel is $2\pi\sigma^2$. In order for this to equal
the area of a rectangular grid cell, and thus lead to the same number of trials as
if the kernel had the shape of a grid cell, the grid constant has to be
$\sqrt{2\pi}\,\sigma$}, to minimize the
correlations between the significances. In practice, a finer
grid may be chosen if the oversampling correlations are taken into account,
which \breftwo{is} not pursued here for clarity. The exposure ratios
$a_{\w}^m$ are considered to be on-time ratios, and the total number of generated
background events are $80000$. For each of the simulations, the
following plots are provided, both for in the presence and absence of a source in the field of view:

\begin{itemize}
\item Measured distributions of events ($N^{\w,m}_i$).
\item Kernels $g^{\w}_i$ and $g^{m}_i$ (\eref{eq:avkern}), assuming the coordinates
of the source.
\item Likelihood as a function of $\phi$, assuming again the coordinates 
of the source, to show how well $\phi\tin{sup}$ is defined.
\item Resulting significance skymap and distribution.
\end{itemize}

The root-finding was done with the bisection method, and no
other approximations were made, although the fact that most summands in
\eref{eq:lllderphi} are almost zero may easily be exploited to build a more
efficient algorithm.

\subsection{Case 1: W=2, M=1, asymmetric exposure shape, equal on-time,
source at observation center} 

\begin{figure}
\centering
\includegraphics[width=12cm]{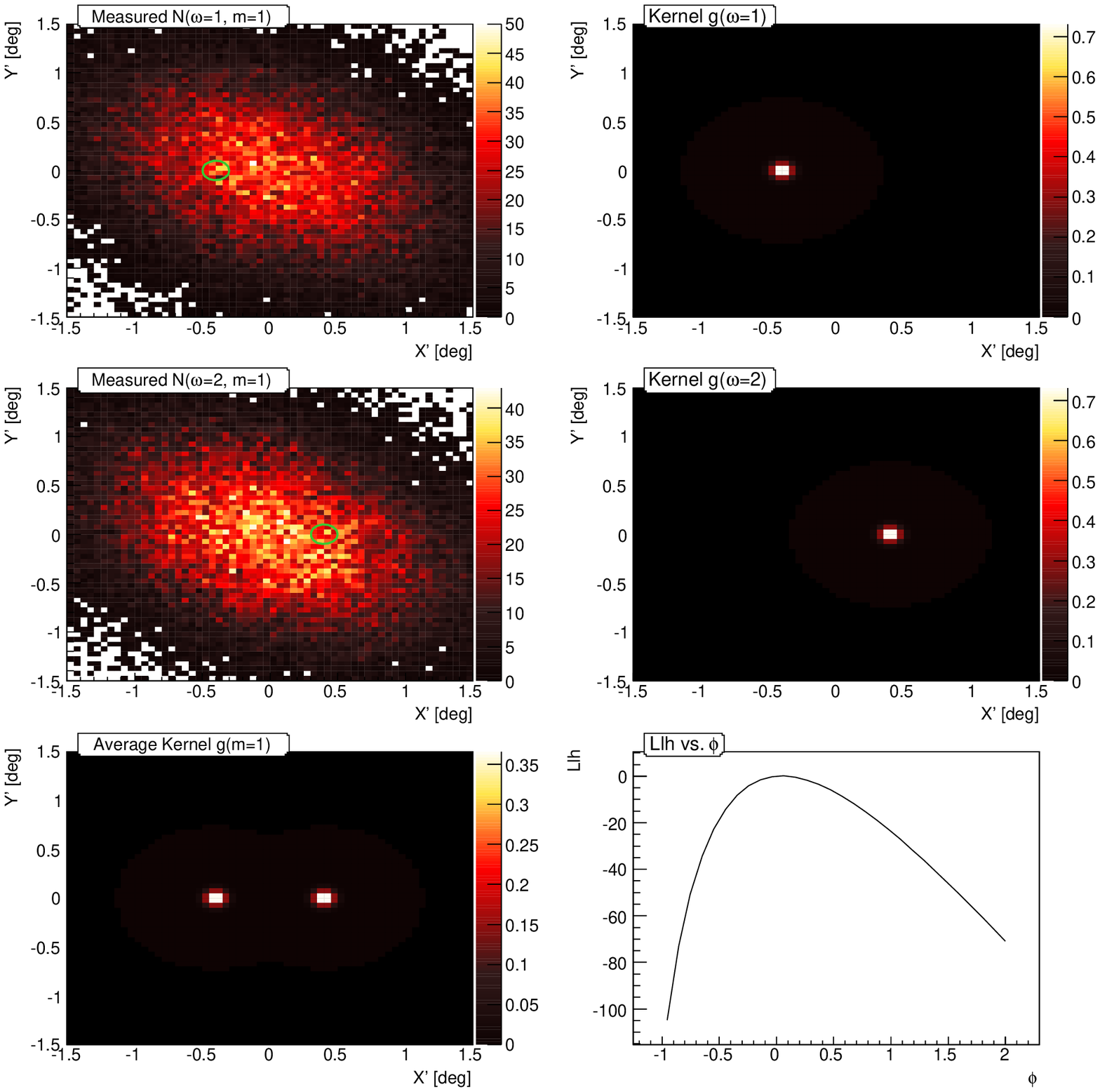}
\includegraphics[width=12cm]{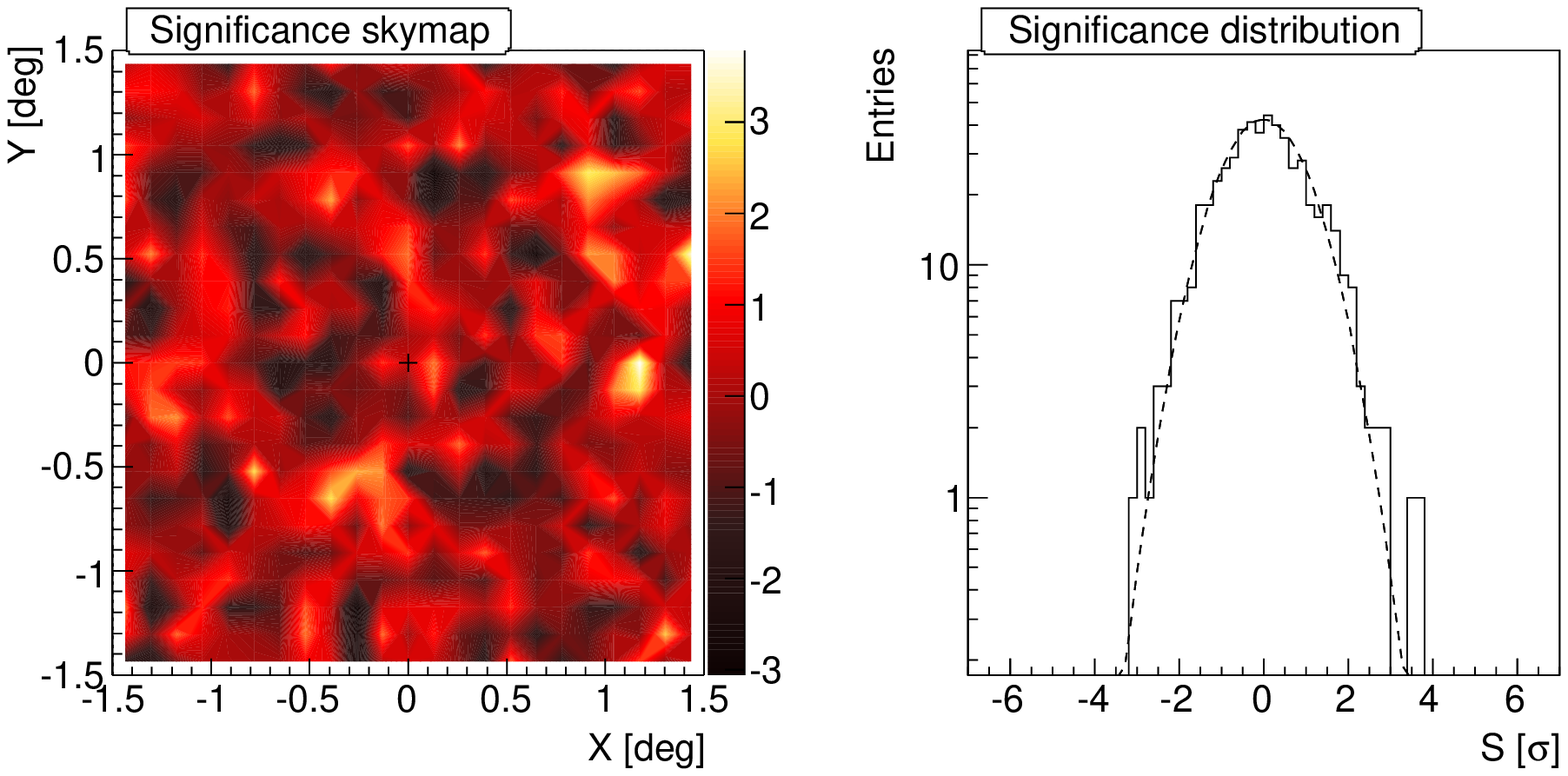}
\caption{
 Case 1: W=2, M=1, asymmetric exposure shape, equal on-time, no source. In the
upper left, the simulated measured distributions are shown, in coordinates relative to
the observation center. Right of it are
the kernels $g^{\w}_i$, assuming the source to be at the observation center, and below is the corresponding average kernel
$g^{m}_i$. The graph in the third row on the right is the log-likelihood as a function of $\phi$, and in the bottom row, the skymap and significance distribution are
shown (the dotted line is a Gaussian of $\mu = 0$ and $\sigma = 1$). The observation center is marked with a cross in the sky map, and green
circles in the data histograms.
}
\label{fig:case1_0}
\end{figure}

This is, for an asymmetric exposure, a simple, ideal case. The source is at
$(0/0)$ in absolute sky coordinates, and the two wobble positions are offset
by $\pm 0.4\dg$ in $X$-direction. Figure~\ref{fig:case1_0} shows the results.
No significant excess can be
found and the null-hypothesis
distribution is compatible with a Gaussian with a mean of zero and with a width of $\sigma=1$. 

Despite the asymmetric exposure, $S\tin{LM}$
can be evaluated correctly, summing up the off-events for a given on-region in the
same region of the opposite wobble set. As mentioned
in~\ref{sec:lima}, the radius inside
which events are integrated is basically a free parameter\footnote{There are
recipes to determine which integration radius is appropriate, but a
discussion or comparison of those is not within scope of this paper.}, so
the significance \breftwo{is scanned in a range of} radii between $1-3\,\sigma\tin{PSF}$,
to make sure that the comparison is not artificially biased \bref{in the favour of
the new formulae presented here}. In the
absence of a signal (Fig.~\ref{fig:case1_0}), the significance \breftwo{found} at
$(0/0)$ is $0.4\sig$ with
\eref{eq:significance} and $S\tin{LM} = 1.1\sig$.

\begin{figure}
\centering
\includegraphics[width=12cm]{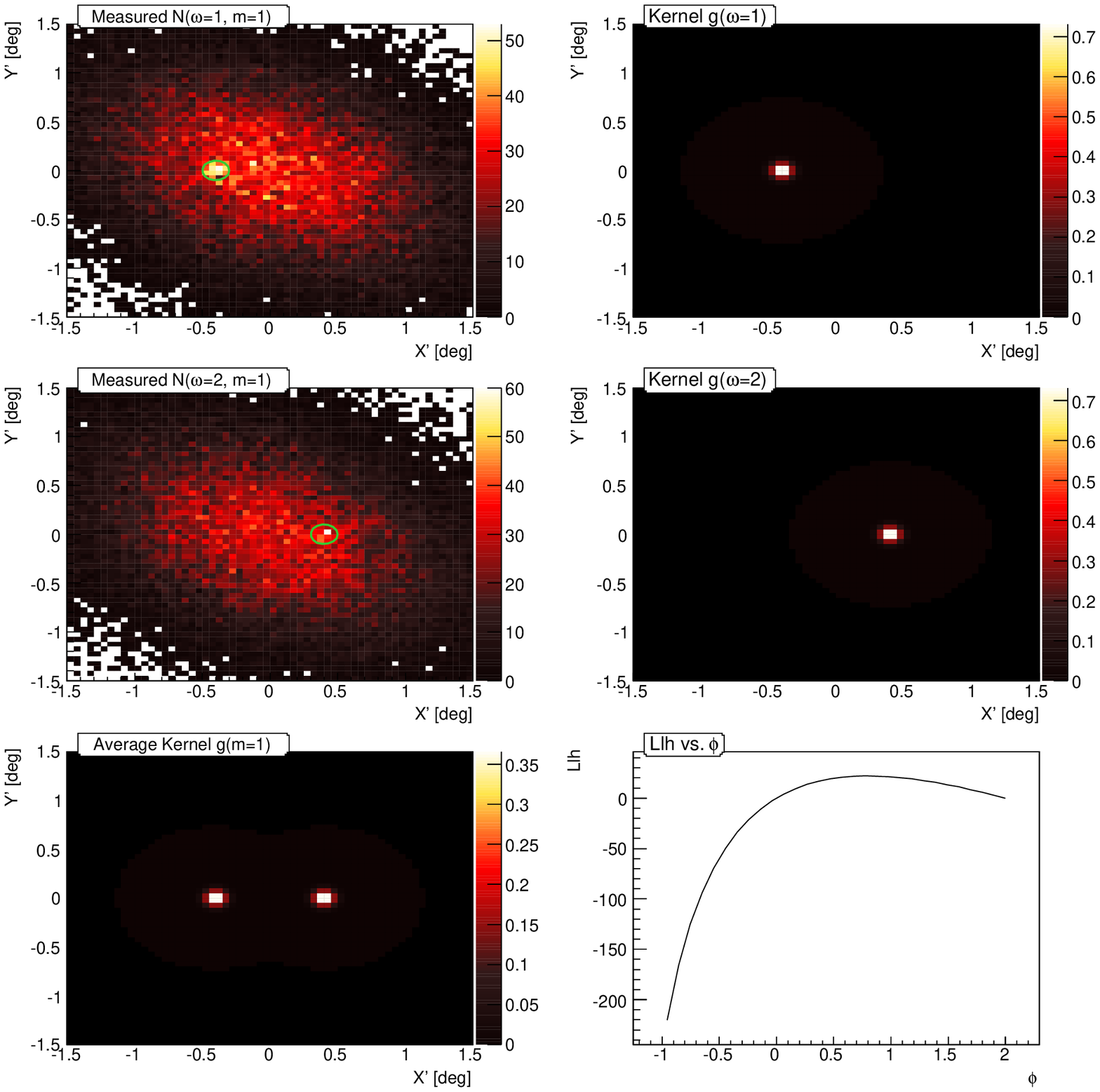}
\includegraphics[width=12cm]{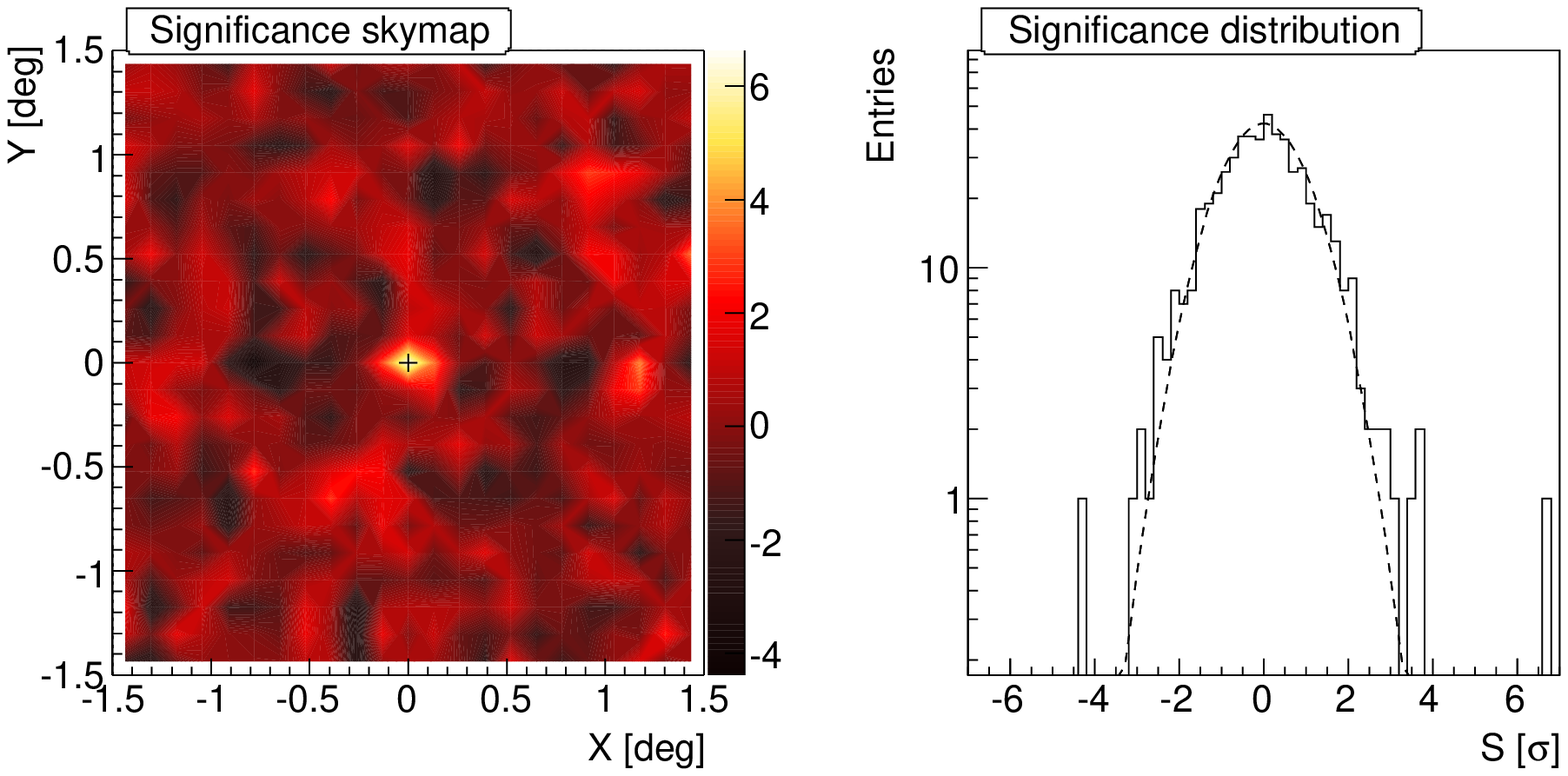}
\caption{
 Same as Fig.~\ref{fig:case1_0}, but with 300 excess events introduced at the
observation center.
}
\label{fig:case1_250}
\end{figure}

Figure~\ref{fig:case1_250} is the same scenario, but inserting 300 excess events
at the observation center. The excess is detected at
$6.6\sig$ with \eref{eq:significance} and  $S\tin{LM}=6.3\sig$. Since
the source happens to be exactly on a grid point (at $(0/0)$), basically only
one grid point shows the signal, proving that the correlations between the
significances are negligible. In the significance distribution, this
one value peaks out while
not disturbing the general Gaussian appearance of the bulk of entries. As negative $\phi$ are allowed (in order to see the negative
part of the distribution), the ambiguity of excess and lacking acceptance
causes minor downward artefacts at $(0.0/\pm0.8)$ in the skymap and at the
negative end of the
significance distribution. This is an expected feature of the algorithm (see
Sec.~\ref{sec:caveats}).

\begin{figure}
\centering
\includegraphics[width=12cm]{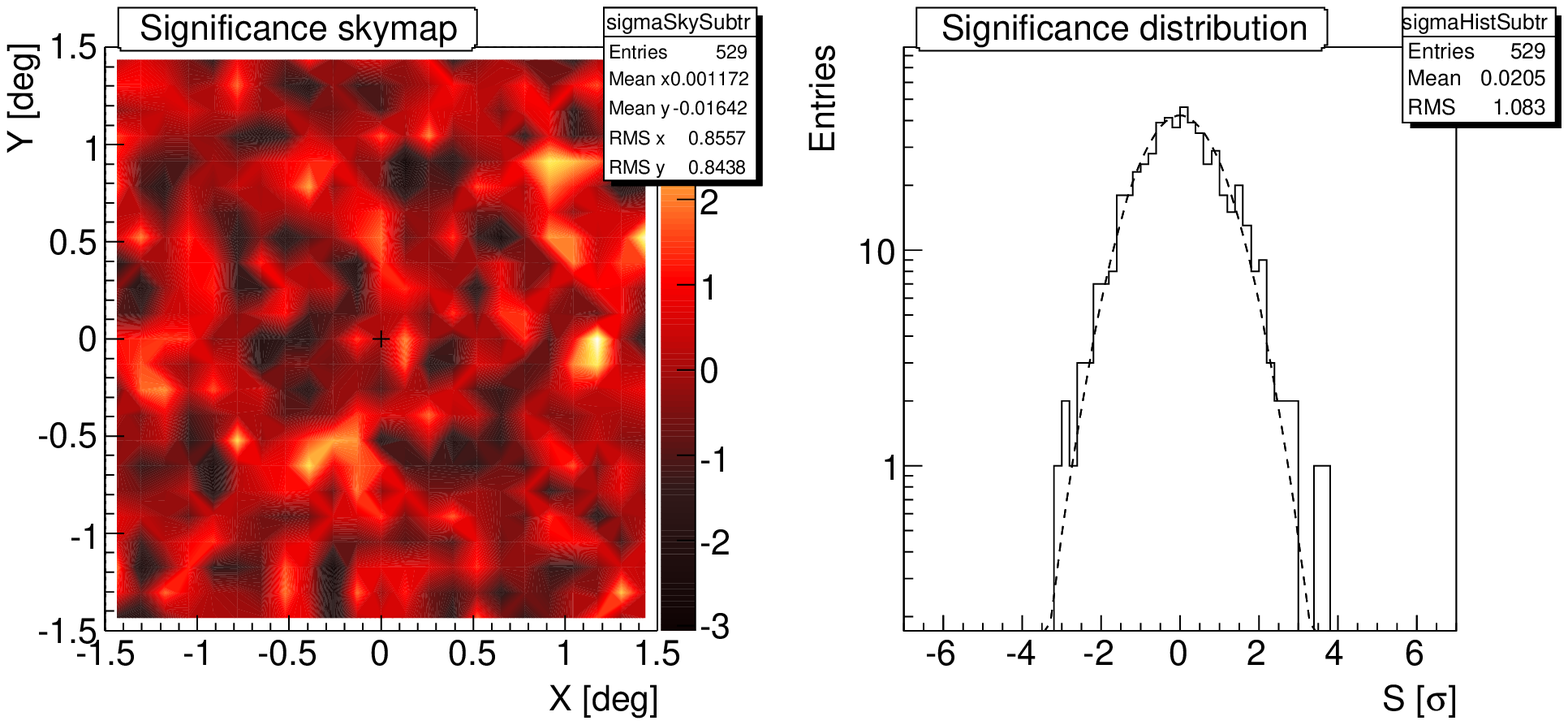}
\caption{
 Same as Fig.~\ref{fig:case1_250}, but with the detected excess included as an
established source in the null hypothesis. The significance map is again
consistent with the null hypothesis. Consequently, the log-likelihood function
peaks at a \breftwo{relative excess} parameter ($\phi$) that is significantly larger than zero.
}
\label{fig:case1_250_subtract}
\end{figure}

Finally, Fig.~\ref{fig:case1_250_subtract} is the same setup as before, but
the detected source is included in the null hypothesis  to validate the
detection and look for possible other sources (see
Sec.~\ref{sec:multisource}). The significance map is again
empty, the downward artefacts are gone and the new null hypothesis is met.

\subsection{Case 2: W=4, M=2, pointing-dependent exposure shape, very different
exposure times, one additional off-data set, extended source with large
offset}\label{subsec:case2} 

\begin{figure}
\centering
\includegraphics[width=12cm]{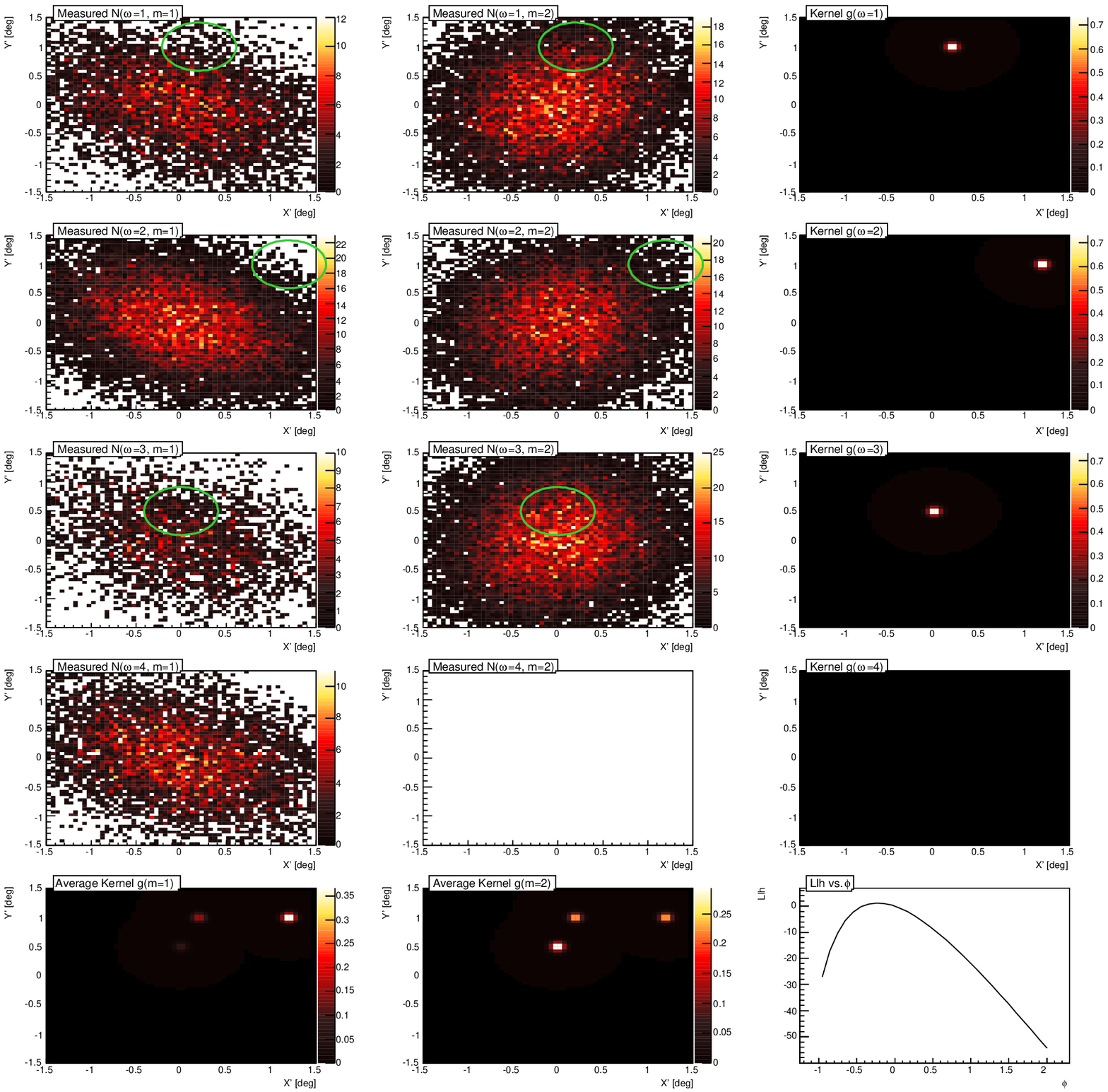}
\includegraphics[width=12cm]{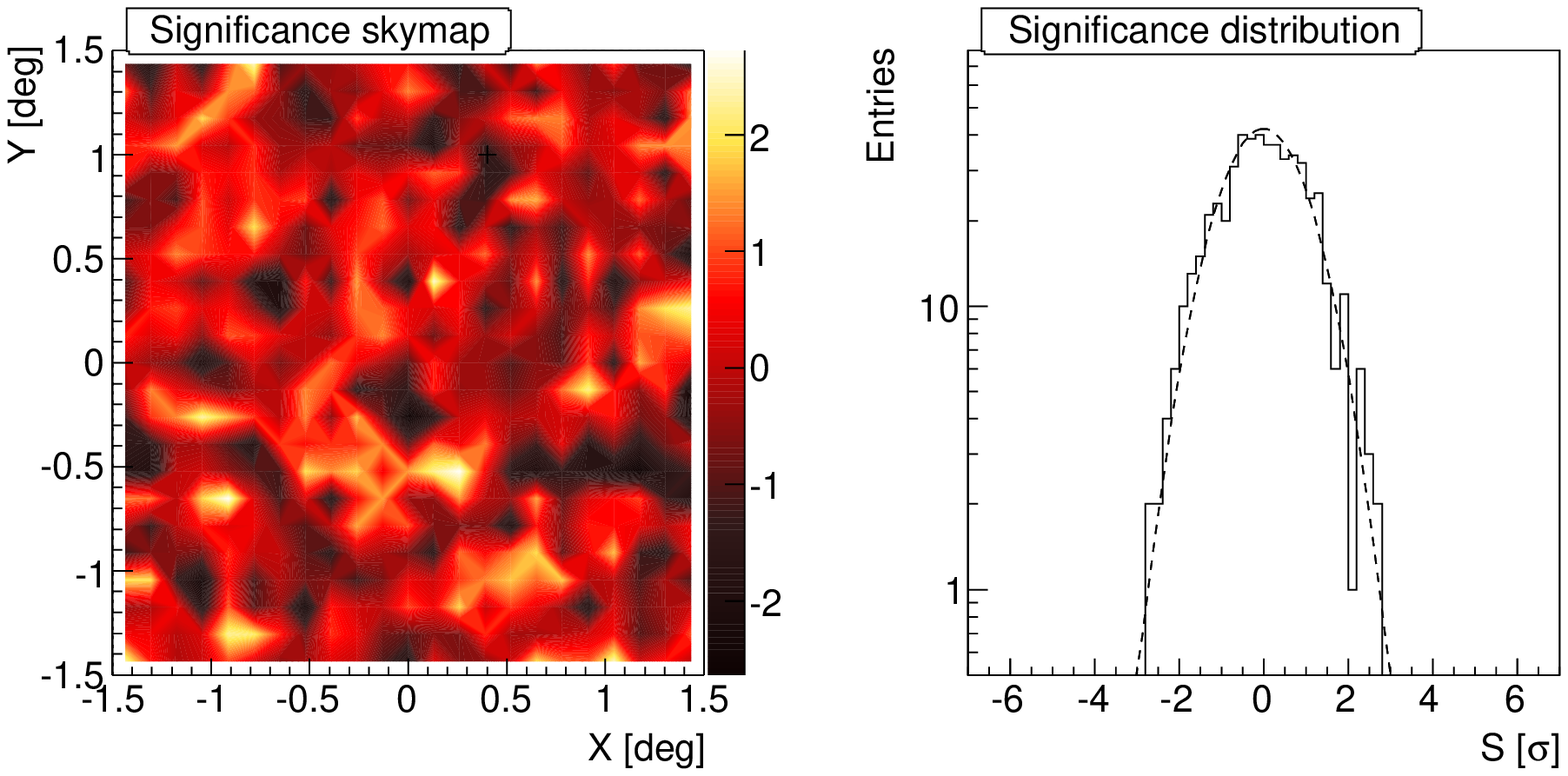}
\caption{
 Case 2: W=4, M=2, pointing-dependent exposure shape, very different
exposure times, one additional off-data set, 
The example kernel plots refer to
the sky position $(0.4/1.0)$, marked with a cross in the significance skymap, where a source is injected in
Fig.~\ref{fig:case2_1200}.
}
\label{fig:case2_0}
\end{figure}

This is a very complex scenario that benchmarks the flexibility of the test
statistic and the 
Likelihood Ratio Skymapping technique. The background events are
distributed over 7 sub-histograms consisting of 3 wobble sets taken in two
different \bref{operating conditions}, and an additional set of off-data for one of the
\bref{operating conditions}. As can be seen in Fig.~\ref{fig:case2_0}, the data can be
combined without any problem, yielding an empty skymap and a Gaussian
null-hypothesis distribution.

\begin{figure}
\centering
\includegraphics[width=12cm]{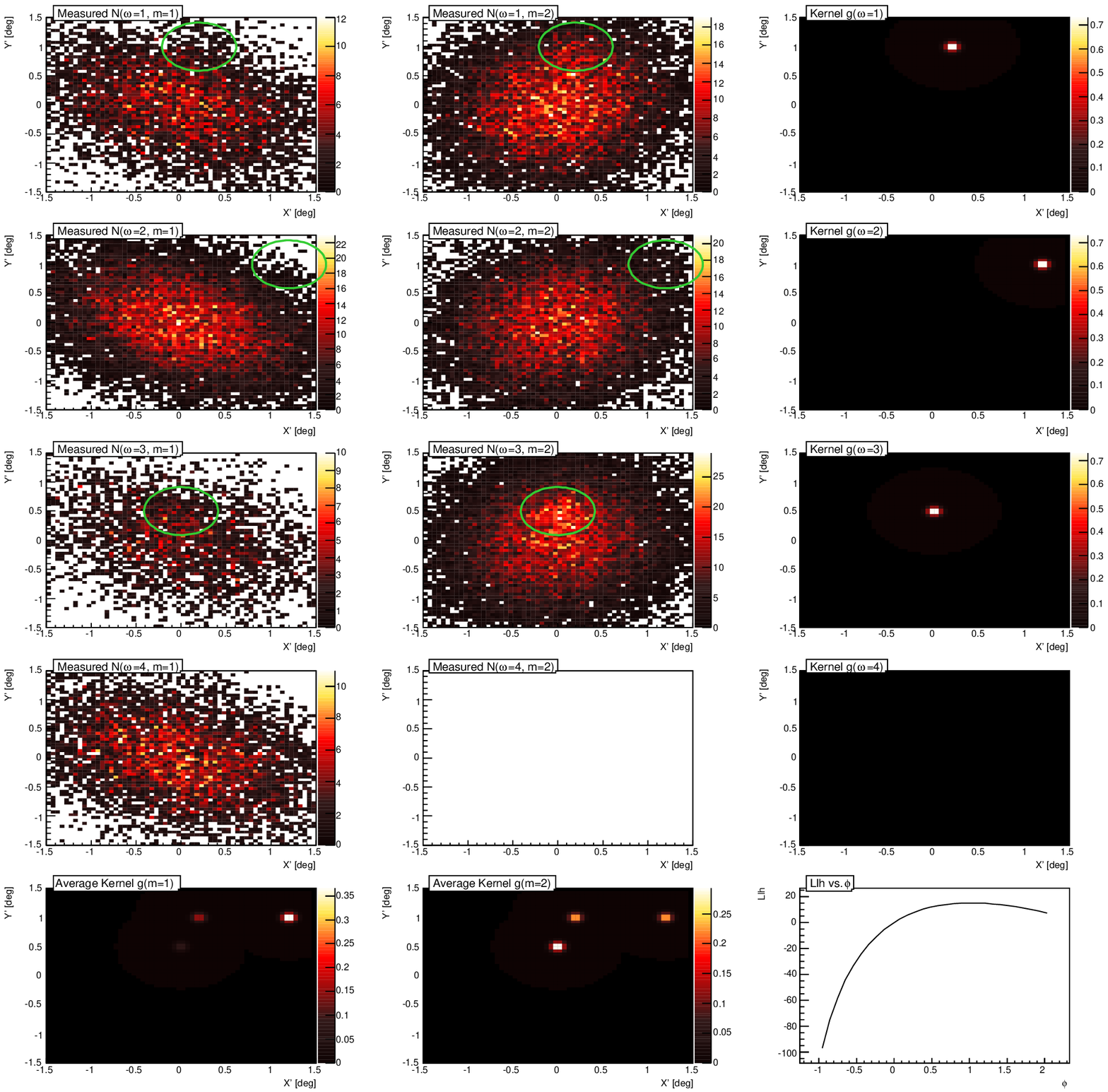}
\includegraphics[width=12cm]{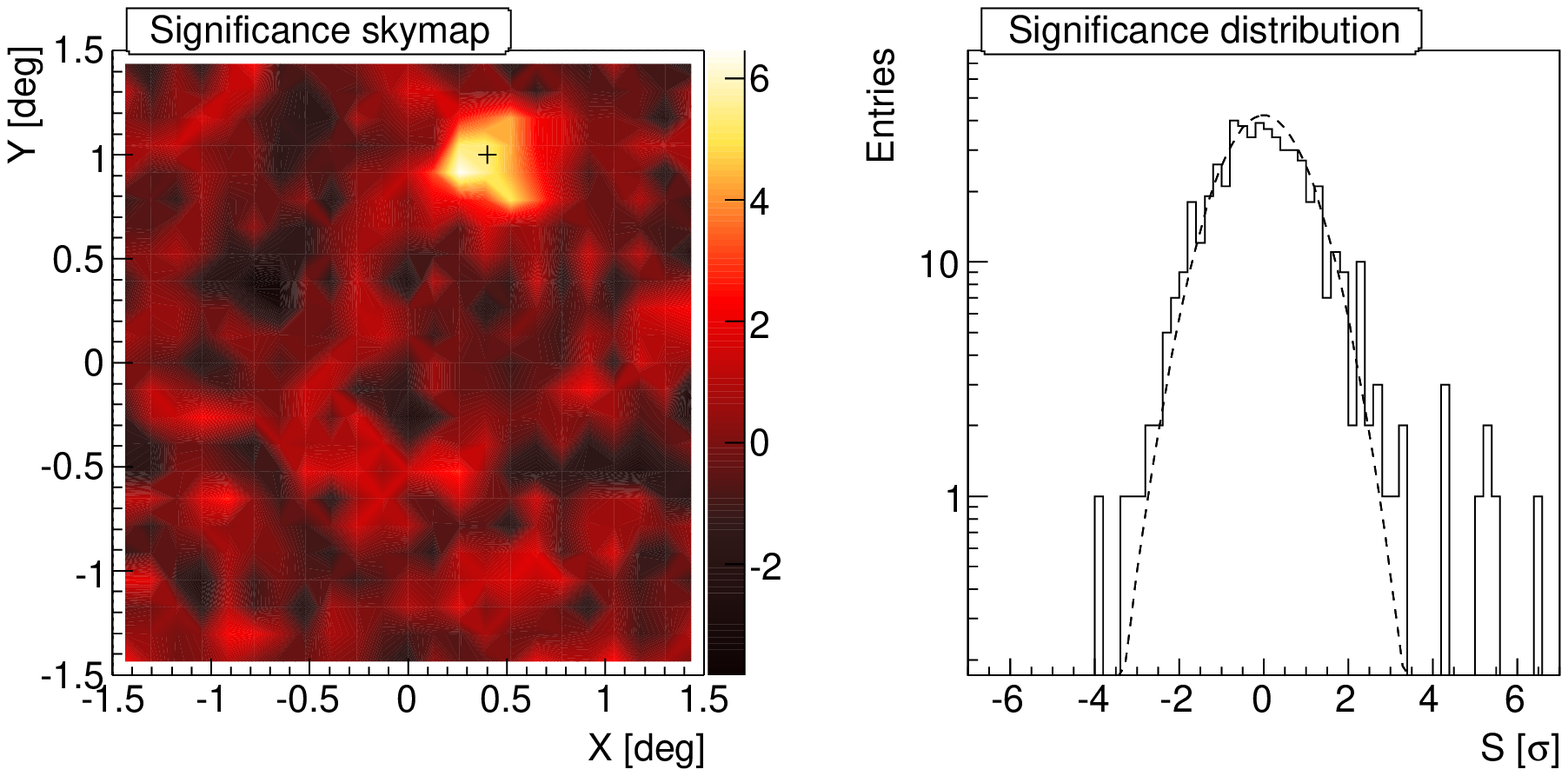}
\caption{
 Same as Fig.~\ref{fig:case2_0}, but with 1200 excess events introduced at the
position marked with a green circle in the upper histograms.
}
\label{fig:case2_1200}
\end{figure}

Figure~\ref{fig:case2_1200} is the same scenario, but with 1200 photons
added to mimic an extended source ($\sigma\tin{src} = 0.2\dg$) at the
large-offset sky
coordinate $(0.4/1.0)$. Even using the same
PSF-Kernel as before, the source can be detected ($5.5\sig$).

\subsection{Gaussian property of \eref{eq:significance} and  comparison to
$S\tin{LM}$} \label{subsec:gauslima}

\begin{figure}
\centering
\includegraphics[width=8cm]{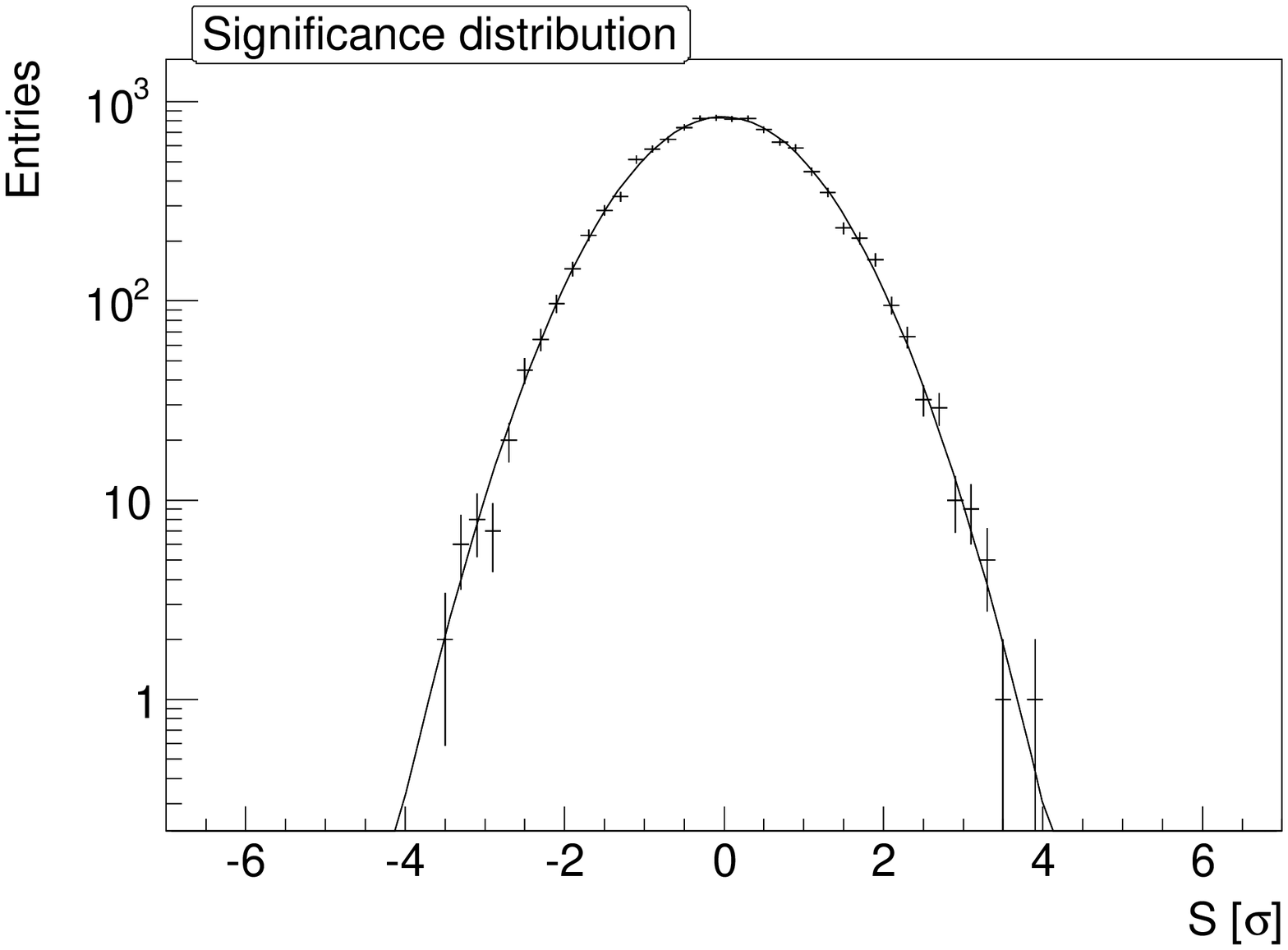}
\caption{
Accumulated significance distribution of 20 simulated skymaps as in Sec.~\ref{subsec:case2}.
Despite the complex setup of several wobble \bref{positions and operating conditions}, and the
skymapping caveats outlined in Sec.~\ref{sec:caveats}, the significance
distribution follows very well the shape of a Gaussian distribution (the solid
line is the Gaussian fit to the data discussed in the text).
}
\label{fig:largestat}
\end{figure}

Figure \ref{fig:largestat} shows a simulation of significances calculated for
the whole field of view of 20 signal-free skymaps simulated as in case 2
(Sec.~\ref{subsec:case2}).
The distribution contains also bins of comparably small exposure and bins for which
the optimization of the $\phi$-parameter reached its boundary condition. No
treatment of the remaining correlation between the significances is
undertaken. 
A Poissonian likelihood fit to the distribution yields Gaussian parameters of
$\mu = -0.012 \pm 0.010$ and $\sigma = 1.006 \pm 0.007$. The $\chi^2 = 40.5$ 
may be compared to the 37 non-zero bins of the distribution. This verifies
both \eref{eq:significance} as a Gaussian significance and the suggested
recipes to deal with its features in the skymapping case.

\begin{figure}
\centering
\includegraphics[width=12cm]{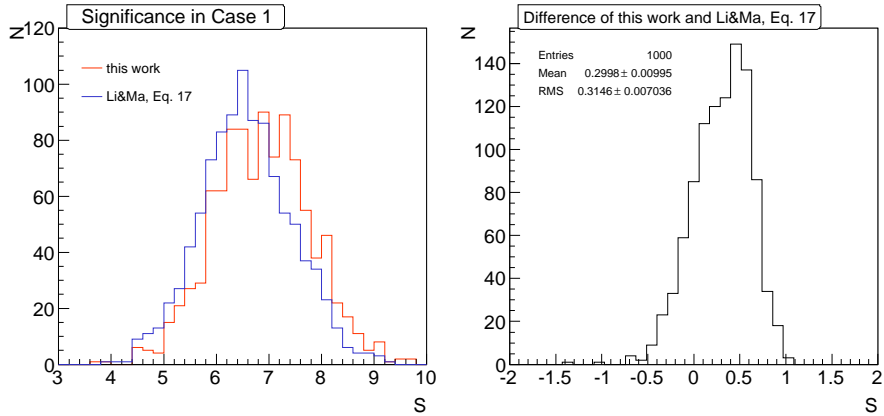}
\caption{
 Comparison of significances obtained with \eref{eq:significance} and
$S\tin{LM}$ (with optimized $\theta^2$-cut). Despite the trials implied by
the optimized $\theta^2$-cut, \eref{eq:significance} yields a systematically
higher significance.
}
\label{fig:limacomp}
\end{figure}

In the above case 1, the test statistic of \eref{eq:significance} is
somewhat higher than $S\tin{LM}$, even though
the latter involves an optimization of the applied integration radius (which
is expected to lead to a selection bias). To study this possible trend
systematically, Figure~\ref{fig:limacomp} shows the distribution of many
significances
obtained for the same case simulated 1000 times. It shows that while  \eref{eq:significance}
does not always lead to a higher significance, there is a significant trend
that in average it does.

Repeating this study with a lower background level, however, the effect is
much weaker, while
with a higher background density, the effect is stronger. This
is assumably because with a low background density, the optimized
$\theta^2$-cut will be large and include the whole signal, reducing the
advantage of the known kernel from which \eref{eq:significance} benefits.

\breftwo{The conclusion can be drawn} that the test statistic presented in this work performs
slightly better in cases with low
signal-to-background ratios, and similar in cases where the dominant
statistical uncertainty is that of the signal itself.

\section{Conclusions}

This paper derives a new generalized test statistic (\eref{eq:significance}) that can be used for
significance calculation in VHE astronomy. The advantages over the
existing test statistics are that it flexibly takes into account any number of
data subsets from different wobble coordinates and \bref{operating conditions of
the system}, even if
the acceptance shape is
very irregular \bref{and different between these operating conditions}. Also, it
\bref{takes}
advantage of a known gamma-ray PSF while not requiring the optimization of an
integration radius ("$\theta^2$-cut").

The test statistic can be applied to any position in the field of view, so it is very suitable
for skymapping purposes. The advantages of this approach is that the test
statistic only makes minimal assumptions on the acceptance field of view and
does not require any exposure symmetry or Monte Carlo simulations. It is hence
unaffected by many systematic uncertainties. \bref{A} "Likelihood Ratio
Skymapping" procedure \bref{is suggested} in Sec.~\ref{sec:procedure}.

The log-likelihood function (\eref{eq:lllhfphi}) can also be applied to fit the
shape and position parameters of the source. \bref{The} formulae are furthermore
extendable to accomodate established sources in the field of view, a
non-homogeneous PSF shape or gamma-to-hadron acceptance ratio, an unbinned
analysis approach or the "orbit" observation mode. \bref{If the background
event density is sufficiently high, the relative} \breftwo{excess} parameter
$\phi$ is well-suited to calculate a gamma-ray flux map of the field of view.

\bref{In several simulated scenarios it is verified} that the test statistic can not only
handle the difficult situations it is designed for, but \bref{also seems to be
systematically higher in the signal case, and therefore more sensitive,}
than the commonly used test statistic after Li and Ma \cite{lima}, Eq.\ 17. 

\section{Acknowledgements}

 The author wishes to thank the MAGIC/CTA groups at IFAE, UAB and ICE for
discussion of the results, and in particular Michele Doro, Daniel Mazin, Abelardo Moralejo,
 Takayuki Saito, Julian Sitarek and Victor Stamatescu for their useful
comments on the draft. Also the support of the "Juan de la Cierva" program of the Spanish
 MICINN is gratefully acknowledged.

\bibliographystyle{elsarticle-num}
\bibliography{llh_skymap}

\end{document}